\documentclass[12pt]{article}
\usepackage{amssymb,amsmath,graphicx,psfrag,url}
\psfrag{E}{$E_0$}
\psfrag{Y}{$Y_0$}
\psfrag{A}{$A$}
\psfrag{D}{$D$}
\psfrag{e}{$\epsilon$}
\psfrag{W}{$W$}
\psfrag{es            }{$\partial \epsilon/\partial\ln{\epsilon_0}$}
\psfrag{ys               }{$\vert\partial \epsilon/\partial\ln{Y_c}\vert$}
\psfrag{ew              }{$\partial W/\partial\ln{\epsilon_0}$}
\psfrag{yw                }{$\vert\partial W/\partial\ln{Y_c}\vert$}
\title{Quantifying ``Cliffs'' in Design Space}
\author{J. I. Katz\thanks{email:{\tt katz@wuphys.wustl.edu}, Tel: 314-935-6202, Facs: 314-935-6219}\\ Los Alamos National Laboratory \\ Los Alamos, 
N.~Mex.~87545 \and Lawrence Livermore National Laboratory \\ Livermore, 
Cal.~94550 \\ \and Department of Physics and \and McDonnell Center for the
Space Sciences \\ Washington University \\ St. Louis, Mo.~63130}
\date{\today}
\begin{document}
\maketitle
\begin{abstract}
Purpose: {\bf This paper studies the regions of parameter
space of engineering design in which performance is sensitive to design
parameters.  Some of these parameters (for example, the dimensions and
compositions of components) constitute the design, but others are intrinsic
properties of materials or Nature.  The paper is concerned with narrow
regions of parameter space, ``cliffs'', in which performance (some measure
of the final state of a system, such as ignition or non-ignition of a
flammable gas, or failure or non-failure of a ductile  material subject to
tension) is a sensitive function of the parameters.  In these regions
performance is also sensitive to uncertainties in the parameters.  This is
particularly important for intrinsically indeterminate systems, those whose
performance is not predictable from measured initial conditions and is not
reproducible.}

Design/methodology/approach: {\bf We develop models of ignition of a flammable
mixture and of failure in plastic flow under tension.  We identify and
quantify cliffs in performance as functions of the design parameters.  These
cliffs are characterized by large partial derivatives of performance
parameters with respect to the design parameters and with respect to the
uncertainties in the model.  We calculate and quantify the consequences of
small random variations in the parameters of indeterminate systems.}

Findings: {\bf We find two qualitatively different classes of performance cliffs.
In one class, performance is a sensitive function of the parameters in a
narrow range that separates wider ranges in which it is insensitive.  In the
other class, the final state is not defined for parameter values outside
some range, and performance is a sensitive function of the parameters as
they approach their limiting values.  We find that sensitivity of
performance to control (design) parameters implies that it is also sensitive
to other parameters, some of which may not be known, and to uncertainties of
the initial state that are not under the control of the designer.  Near or
on a cliff performance is degraded.  It is also less predictable and less
reproducible.}

Practical implications: {\bf Frequently, design optimization or cost minimization
leads to choices of engineering design parameters near cliffs.  The
sensitivity of performance to uncertainty that we find in those regimes
implies that caution and extensive empirical experience are required to
assure reliable functioning.  Because cliffs are defined as behavior on
the threshold of failure, this is a reflection of the tradeoff between
optimization and margin of safety, and implies the importance of ensuring
that margins and uncertainties are quantified.  The implications extend far
beyond the model systems we consider to engineering systems in general.}

Originality/value: {\bf Many of these considerations have been part of the 
informal culture of engineering design, but they were not formalized until
the methodology of ``Quantification of Margins and Uncertainty'' was 
developed in recent years.  Although this methodology has been widely
used and discussed, it has only been published in a small number of
reports (cited here), and never in a journal article or book.  This paper
may be its first formal publication, and also its first quantitative
application to and illustration with explicit model problems.}
\end{abstract}


\newpage
\section{Introduction}
Many complex engineering systems are difficult, expensive, impossible or
forbidden to test throughout their full range of required performance (for
example, to destructive failure).  Examples include nuclear reactors,
industrial facilities such as oil refineries or chemical processing plants,
dams and flood control structures, large machines such as power turbines,
ships and aircraft, systems required to have lives longer than the duration
of any feasible test, such as containers for the permanent sequestration of
radioactive waste, and nuclear weapons.

The designer must reconcile the goals of optimizing performance, implicitly
minimizing the use of material, human and financial resources, and of
controlling its uncertainty.  He must be concerned with the margin against
performance out of a specified range, and especially with the margin against
catastrophic failure.

Limitations on experiment and testing force the designer to a heavy reliance
on computation, supplemented by a very small number of tests.  In some cases
this number is zero; perhaps the most famous example was the ``Little Boy''
uranium atomic bomb, which was used in combat on the basis of theory and
laboratory data, but without a full system test (Rhodes 1986).  When design
is heavily dependent on computation it is essential to understand and
constrain the uncertainties of the computational results.  There are always
uncertainties in modeling the underlying processes, sometimes quantitative
in their parameters and sometimes qualitative in the models themselves.  For
example, the rates of chemical processes are known with finite accuracy, but
there are fundamental gaps in the understanding of turbulent flow.
Numerical calculation introduces additional uncertainties.
\section{Review of Literature}
The method of Quantification of Margins and Uncertainties (QMU),
developed by Goodwin and Juzaitis 2003 and further explicated by Eardley,
{\it et al.} 2005 and by Pilch, Trucano and Helton 2006, formalizes
these issues.  This literature established the framework, but did not 
quantify it in specific model (or real-world) problems.  In this paper I
quantify the ``cliffs'' in design parameter space that underlie QMU
through the use of illustrative ``toy'' models that are simple
enough for quantitative analysis but that show the qualitative features of
real design problems.  In particular, I discuss the issues that must be
addressed if the method of QMU is to be applied in practice, illustrate
the sensitivity of performance both to design parameters and to other
uncertainties, and consider the issue of intrinsic indeterminacy.

In this paper we consider two ``toy'' problems that illustrate the
phenomenon of cliffs.  The first is the ignition of a flammable gas
mixture by a small spark.  This is a simple model of the more complex
problem of ignition of a laser fusion capsule (Chang, {\it et al.\/} 2010;
Lindl, {\it et al.\/} 2011; Haan, {\it et al.\/} 2011; Edwards, {\it et
al.\/} 2011).  The second is the well-known problem of plastic failure of a
ductile material in tension (Ugural and Fenster 2011).
\section{Design Parameters}
The prudent designer chooses regions of design space in which the
unavoidable modeling uncertainties imply small performance uncertainties,
and avoids, if possible, regions in which they imply large performance
uncertainties.  We describe the design by figures of merit 
\begin{equation}
Y_k \left(\{x_i\},\{p_j\}\right).
\end{equation}

The $\{x_i\}$ are the uncertain parameters of the processes involved.  For
example, they may be chemical rate coefficients, material properties,
parameters of turbulence models or nuclear cross-sections.  Some of these
can be measured (to finite accuracy), but are not under the control of the 
designer.  Even the existence of others may not be appreciated.

The $\{p_j\}$ are design or control parameters, such as those describing
the shape of a wing, the thickness of a structure, an applied force or the
concentration of a chemical reactant.  They are under the control of the
designer, usually to high accuracy.

The designer may not know the uncertainties in the $\{x_i\}$.  Physical
parameters can usually be straightforwardly measured and the uncertainties
in their measurements estimated with some reliability.  In contrast, more
complex models such as those of turbulence and reaction networks have
uncertainties whose magnitude and implications are more difficult to 
constrain.  Hence it may not be possible to establish confidence intervals
of the $Y_k$ by performing a series of calculations in which the $\{x_i\}$
are varied through known ranges of uncertainty. 

The prudent designer will attempt to choose values of the $\{p_j\}$ for
which the uncertainties in the $\{x_i\}$ have minimum, or at least small,
influence on the $Y_k$.  In other words, he will attempt to choose regions
of $\{p_j\}$ space in which the magnitudes of the partial derivatives
\begin{equation}
\label{partials}
\left\vert\partial Y_k\left(\{x_i\},\{p_j\}\right) \over \partial\ln{x_l}
\right\vert
\end{equation}
are small.  In this expression derivatives with respect to the logarithms of
the $x_l$ are used in order to make the results independent of the
dimensions and scales of these parameters.  However, the logarithmic
derivatives of the $Y_k$ with respect to the $x_l$ are less informative
because they can be large when the $Y_k$ themselves are too small for the
design to be useful.  

Often, design optimization requires choice of $\{p_j\}$ for which the
partial derivatives (\ref{partials}) are large, defining a ``cliff'' in
parameter space.  This may be the result of a requirement to minimize mass,
volume, cost or material.  Then careful quantification of the margins and
uncertainties is necessary because large values of these partial derivatives
imply proportionately large uncertainties in performance $Y_k$.  In
addition, as illustrated in the model problems in this paper, when the first
derivatives are large, so generally are higher derivatives, increasing the
sensitivity to finite uncertainty.


The purpose of this paper is to illustrate and illuminate these qualitative
ideas with simplified but quantitative examplars.  As exemplars I consider 
two simple ``toy'' problems.  These are much simplified models of real
problems, but may provide useful insight into more complex real problems if
they show their qualitative features while still being simple enough for 
their behavior to be transparent.

One toy problem is the ignition of a flammable mixture of gases following
the heating (for example, by a spark) of a small region of the mixture,
and may also be thought of as a model of the ignition of a laser fusion
capsule.  The design (or control) parameter is the initial temperature of a 
reacting region of finite size, corresponding to the energy of an igniting
spark or laser pulse.  The heat of combustion accelerates the reaction, but
conduction carries heat away and diffusion dilutes the reactants.  The
$\{x_i\}$ consist of the parameters describing the reaction rate, $\{p_j\}$
is the initial temperature (equivalent to the spark energy), and $Y$ is the
cumulative energy release.

The second toy problem is the plastic flow, and ultimate failure, of a
work-hardening ductile material under quasi-static tensile load, such as
found in a tensile test machine.  The design (or control) parameter is the
applied tensile force.  As the test sample stretches, it narrows (increasing
the stress) but also hardens, increasing its resistance to plastic flow.
The $\{x_i\}$ are the parameters of the work-hardening model, $\{p_j\}$
is the applied load, and the $Y_k$ are $\epsilon$, the total longitudinal
plastic strain, and the cumulative plastic work per unit volume $W$.  A
practical application is the use of the plastic flow (typically in bending
or crumpling) of ductile elements to dissipate the kinetic energy of a
vehicular collision as plastic work; it is desired to maximize the
dissipation, but the material must not break.
\section{Determinate and Indeterminate Systems}
Some engineering systems are robust against uncertainty: small deviations
from nominal conditions or properties produce proportional deviations in 
performance.  Others are non-robust: performance is so sensitive to small
deviations that it is unpredictable.  The distinction between robust and
non-robust behavior in a determinate system is quantitative, but the
distinction between determinate and indeterminate systems is
qualitative.

An example of a determinate system is the plastic failure of ductile
materials.  If they are stressed beyond their elastic limit (which cannot be
known exactly) by a small amount, the overstress is accommodated by plastic
flow and work hardening, with irreversible microscopic damage but without
catastrophic failure.  This robust behavior is predictable with finite,
usually small, and controlled uncertainty.

If a ductile material is subject to a larger overstress its behavior may
remain determinate, but not robust: The same test, repeated with slightly
varying conditions, will produce results that are sensitive to those 
conditions (so that it is not robust) but that is predictable if the 
conditions are accurately known.  A very ductile metal may be drawn into a
wire whose length is a rapidly varying, but determinate, function of the
drawing force.  A small overstress produces a small plastic deformation but
a larger overstress, carefully modulated as a function of the resulting
strain, draws an ingot out into a fine wire whose length and diameter are
sensitive functions of the control parameters $\{p_j\}$ (in this example the
$\{p_j\}$ describe the dependence of the drawing force on the extension).

These issues are particularly important if the system is not determinate.
In such a case, even when test data are available, they may have little
predictive value.  A single test of a determinate system establishes its
performance to the accuracy and reliability (which must include the 
possibility of human error) of the test.  For a determinate system
application to other exemplars of the same design requires consideration of
variations in the initial conditions, but these usually can be measured
quite accurately.

This is not true for an indeterminate system, whose full distribution of
outcomes can only be determined statistically, and generally only from a
large body of data.  A familiar example is brittle failure.  The degree of
indeterminicity may be quantified by its Weibull modulus (Weibull 1951,
Freudenthal 1968); although the behavior of a single specimen is
indeterminate, it is bounded.  In order to establish a $100 p$\% confidence
interval of the range of outcomes it is necessary to perform ${\cal O}
(1/(1-p))$ tests.  This is typically a few times $1/(1-p)$, the
multiplicative factor depending on the confidence required in the {\it
limits\/} of that interval.  Very often, this is not feasible; determining a
95\% confidence interval requires $\gg 20$, perhaps 50--100, tests.

Indeterminate systems may be the result of intrinsically statistical
processes, such as quantum mechanical measurement.  They may also be the
result of exponential growth of imperfections (such as internal defects,
heterogeneities, surface scratches and deviations from nominal surface
finish or configuration) in initial conditions that cannot be reduced to the
exponentially fine accuracy that would be required for a determinate
calculation.  In other cases, particularly those involving turbulent flow,
determinate calculation is not computationally feasible.  

It is often not known if a system is determinate, which adds another
source of uncertainty to the interpretation of test data.  Even in an 
indeterminate system the range of possible outcomes is bounded.  These
bounds may be narrow, except near a cliff where they are likely
to be broad.  This is an additional reason why it is important to know
where cliffs exist in design space, to quantify their steepness, and to
avoid these regions.
\section{A Determinate Model: Ignition}
\label{ignite}
A classic example of a phenomenon showing a performance ``cliff'' is the
ignition of a flammable mixture of gases.  It generally requires a minimum
spark energy.  Here I discuss a minimal model of ignition, simple enough for
intuitive understanding, and its quantification by means of the partial
derivatives (\ref{partials}).  It is not meant to be a realistic description
of the actual ignition of flammable gases or an inertial fusion capsule,
although such capsules are a well-known and well-quantified example of a
design problem with a steep cliff (Chang, {\it et al.\/} 2010 Fig.~4).
Because the model is realized in a digital computation it is necessarily
strictly determinate, but it illustrates the sensitivity to parameter
variations characteristic of indeterminate systems.

Energy release is described by the equation of second order kinetics
\begin{equation}
\label{rate}
{dY \over dt} = Y_0 A C^2 \exp{(-E_0/T)},
\end{equation}
where $Y_0$ is the energy of reaction, $A$ is a rate coefficient, $C$ is 
the concentration of each of the two reactants (a stoichiometric mixture is
assumed), $E_0$ is an Arrhenius kinetic barrier to the reaction and $T$ is
the matter temperature in energy units.  Reactions deplete the quantity $Q$
of reactants according to the equation
\begin{equation}
{dQ \over dt} = - {1 \over Y_0} {dY \over dt},
\end{equation}
with the initial condition $Q = Q_0$.  The concentration is described by
\begin{equation}
C = {Q \over V},
\end{equation}
where the reactants are assumed uniformly distributed through a region of
radius $R$ and volume
\begin{equation}
V = {4 \pi \over 3} R^3.
\end{equation}
The temperature is increased by the release of chemical energy according to
\begin{equation}
T = T_0 + {Y \over V}.
\end{equation}

An essential feature of the model is the expansion of the reacting region.
This is assumed to be described by a diffusion equation
\begin{equation}
R = 1 + \sqrt{Dt}
\end{equation} 
with diffusion coefficient $D$.  This reduces both the concentration of
reactants and the temperature.  As a consequence, the model system has two
distinct paths:
\begin{enumerate}
\item For ``low'' values of $T_0$, the reaction rate is low and reactants
diffuse to negligible concentrations before there is any significant energy
release.  Only a small fraction of the reactants ever react; they do not
ignite.
\item For ``high'' values of $T_0$ release of chemical energy accelerates
the reaction rate and most of the reactants are consumed before diffusion
becomes significant.  This corresponds to ignition.
\end{enumerate}
The value of $T_0$ that separates these regimes is determined by the values
of $\{x_i\} = \{E_0, Y_0, A, D, Q_0\}$ ($Q_0$ determines the initial value
of the concentration $C$ by the normalization of the initial radius to
unity).

The two regimes are separated by a ``cliff''.  For values of $T_0$ far from
this cliff, the paths and $\lim\limits_{t \to \infty}Y$ (the total chemical
energy released) are robust and little affected by variations (in this toy
model) or uncertainties (in a quantitative model of a real process) in the 
$\{x_i\}$.  For $T_0$ near the cliff the opposite is true, and the
confidence that can be placed in the path and in $\lim\limits_{t \to 
\infty}Y$, the quantity of interest, is reduced by the uncertainties in the
$\{x_i\}$.  The toy model is useful because these parameters can be varied
at will; in a real-world model even their uncertainties would be imperfectly
known, limiting the confidence that could be placed in the results of any
calculation of behavior near the cliff.  

\begin{figure}
\begin{center}
\includegraphics{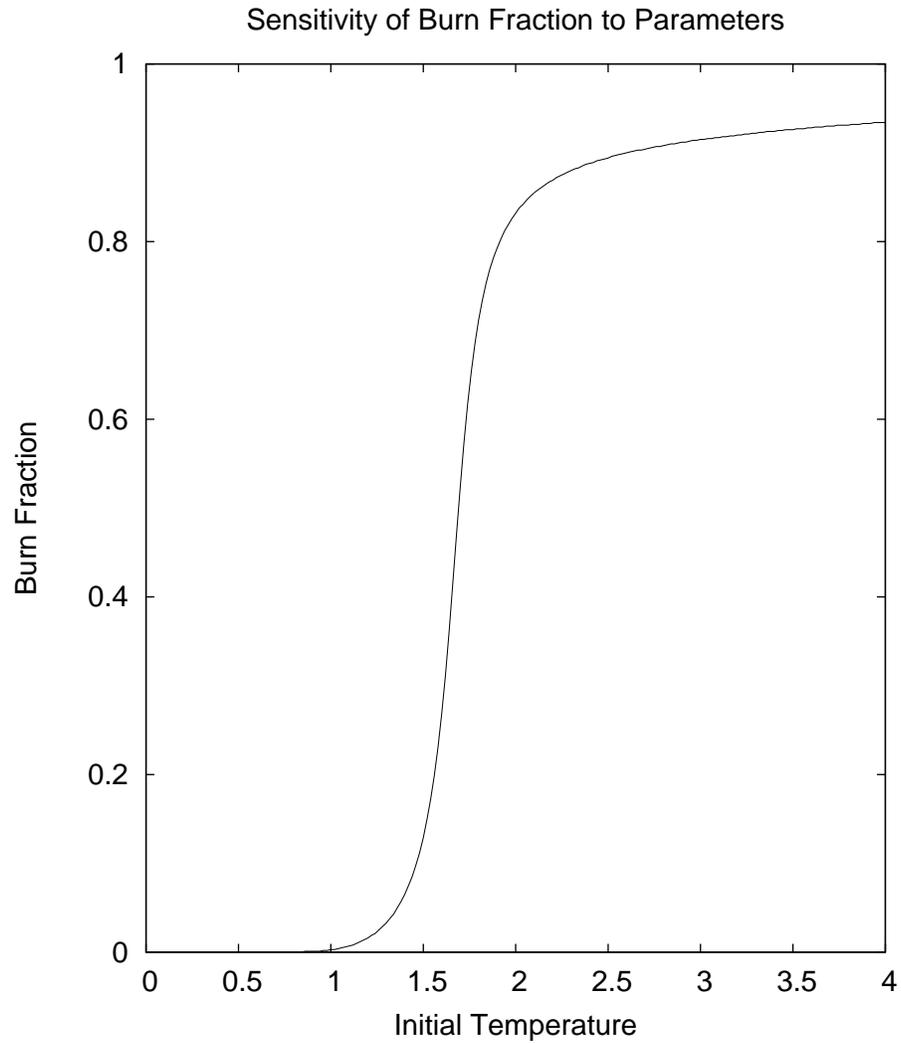}
\end{center}
\caption{Energy release as a function of initial temperature or spark or
laser energy in the toy model.  For $T_0$ significantly below a critical
value the reactants do not ignite and there is negligible energy release,
while significantly above this critical value they ignite and burn nearly to
exhaustion.  These regimes are separated by a cliff at which the energy 
release is a sensitive function of $T_0$.}
\label{cliff}
\end{figure}

For the results shown here $E_0 = 10$, $Y_0 = 30$, $A = 100$, $D = 1$ and 
$Q_0 = 4 \pi/3$ (corresponding to an initial $C = 1$).  Fig.~\ref{cliff}
shows the cumulative energy release $\lim\limits_{t \to \infty}Y$ as a
function of the initial temperature $T_0$.  In the toy model with no spatial
dependence (one spatial zone) $T_0$ is equivalent to an initial spark or 
laser pulse energy.  This behavior illustrates a cliff.

\begin{figure}
\begin{center}
\includegraphics{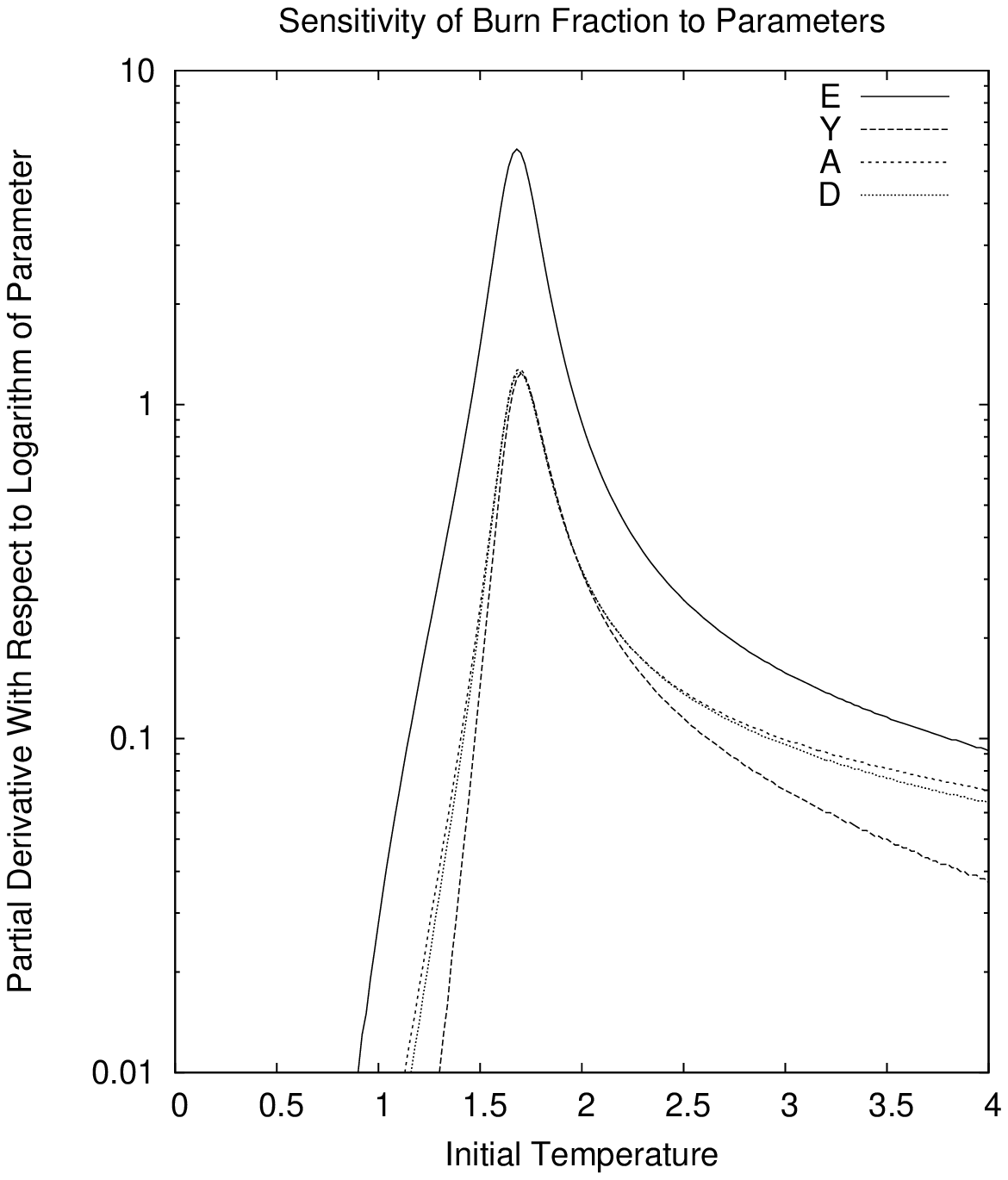}
\end{center}
\caption{Magnitudes of the partial derivatives (\ref{partials}) of the
energy release with respect to the logarithms of the parameters of the 
ignition model.  The partial derivatives have maxima at the location of the
cliff shown in Fig.~\ref{cliff} and indicate the sensitivity of the system
to uncertainties.}
\label{derivs}
\end{figure}

The values of the partial derivatives (\ref{partials}) are shown in
Fig.~\ref{derivs}, normalized to the magnitudes of the corresponding
parameters.  Unsurprisingly, they have narrow peaks at the value of $T_0$
for which a cliff is apparent in Fig.~\ref{cliff}, demonstrating the high
sensitivity of the results to uncertainties in the $\{x_i\}$ for values of
the $\{d_j\}$ corresponding to a cliff. 

\section{A Determinate Model: Necking in Plastic Flow}
Here we consider a very simple model of the narrowing by plastic flow of a
coupon or rod of a work-hardening ductile material under tensile load.   
This describes a familiar quasi-static tension test.  As in the previous
section, we make a one-zone approximation, ignoring any variation of the
necking along the length of the sample.  This is often a good approximation
for a work-hardening material when tension is applied to a long slender
specimen.  We also ignore the elastic strain, both shear and volumetric;
this is generally a excellent approximation for ductile materials that
undergo strains $\gtrsim 0.1$ before failure, because their strength is
typically $\lesssim 10^{-3}$ of their elastic moduli.  Finally, in a
quasi-static test strain rate and work-heating effects are negligible (by
definition).

A tensile force $F$ is applied along the $\hat z$ axis.  The sample has an
initial cross-sectional area $A_0$ and unstrained uniaxial yield strength
$Y_0$.  By conservation of volume in the one-zone model the cross-section is
\begin{equation}
A = {A_0 \over 1 + \epsilon},
\end{equation}
where $\epsilon \ge 0$ is the strain in the direction of the applied
tension.  For the work-hardening law we adopt the empirical form of 
Wilkins and Guinan (1973)
\begin{equation}
Y = Y_0 \left(1 + Y^\prime {\epsilon \over \epsilon + \epsilon_0}\right),
\end{equation}
who find for pure copper $\epsilon_0 = 0.14$ and $Y^\prime = 4$.

\begin{figure}
\begin{center}
\includegraphics{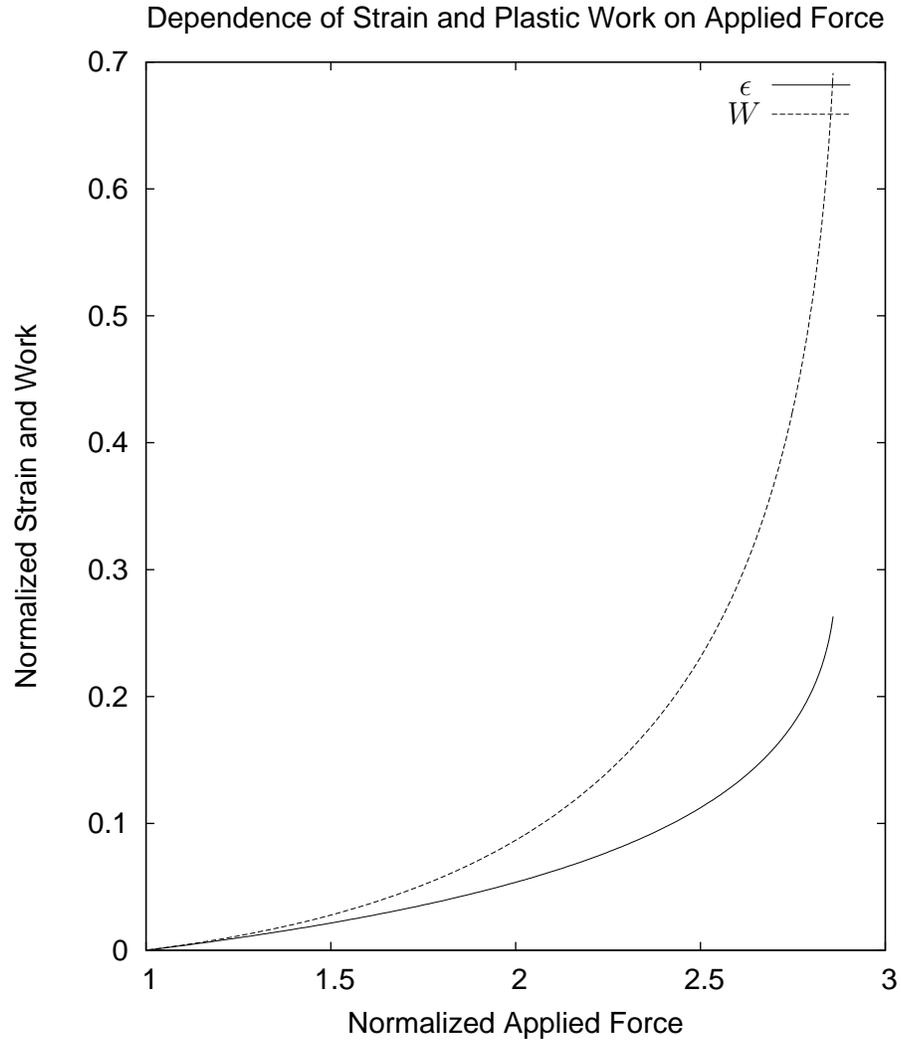}
\end{center}
\caption{\label{epsf} The dependence of longitudinal strain $\epsilon$ and
plastic work $W$ on the applied tensile force $f$ (normalized to the elastic
limit of the unstrained sample), using the empirical parameters for pure
copper found by Wilkins and Guinan (1973).  The curves end at $\epsilon =
\epsilon_{sing} = 0.2632$ at which $f = 2.859$ but $\partial \epsilon/
\partial f$ diverges.  No solutions exist for $\epsilon \ge \epsilon_{sing}$;
the material breaks for $f \ge f(\epsilon_{sing})$.}
\end{figure}

Then the non-dimensionalized force $f$ is given by
\begin{equation}
f \equiv {F \over A_0 Y_0} = {1 \over 1 + \epsilon} + Y^\prime {\epsilon
\over (1 + \epsilon) (\epsilon_0 + \epsilon)}.
\end{equation}
This is a quadratic equation for $\epsilon$ with the solution
\begin{equation}
\epsilon = {[(1 + Y^\prime) - f (1 + \epsilon_0)] - \sqrt{[(1 + Y^\prime) -
f (1 + \epsilon_0)]^2 - 4 f (f - 1) \epsilon_0} \over 2 f}
\end{equation}
shown in Figure~\ref{epsf}.

The derivative
\begin{equation}
{d\epsilon \over df} = {(1 + \epsilon)^2 \over Y^\prime (\epsilon_0 -
\epsilon^2)/(\epsilon_0 + \epsilon)^2 - 1}
\end{equation}
becomes singular, with $d\epsilon / df \to \infty$, at
\begin{equation}
\label{epssing}
\epsilon_{sing} = {- \epsilon_0 + \sqrt{\epsilon_0^2 + \epsilon_0 
(Y^{\prime 2} - 1)} \over Y^\prime + 1} = 0.2632,
\end{equation}
at which $f = 2.859$ and $Y = 3.611 Y_0$.  The singularity corresponds to
failure of the sample; with the assumed work-hardening law, no solutions
exist for $\epsilon \ge \epsilon_{sing}$ or $f \ge f(\epsilon_{sing})$.  The
sample breaks at this value of strain, even though there are no cracks or
stress concentration in the model.

\begin{figure}
\begin{center}
\includegraphics{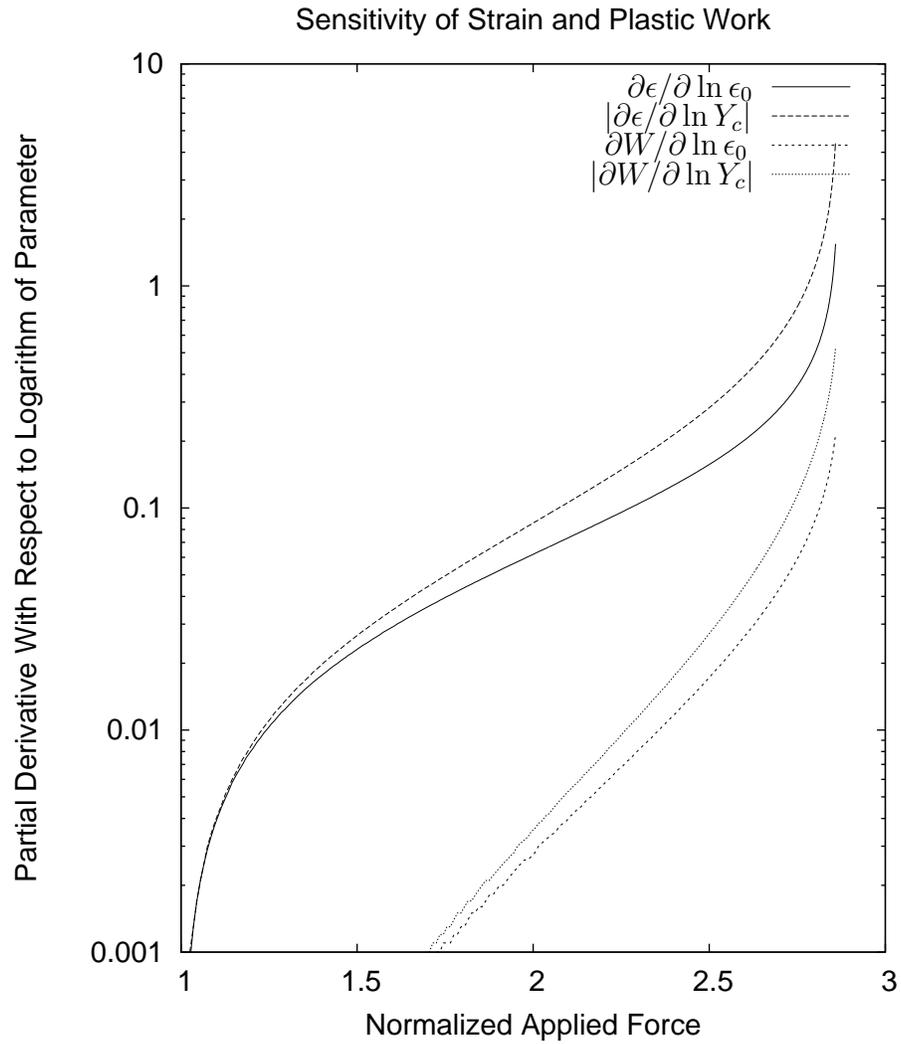}
\end{center}
\caption{\label{epsderiv} Magnitudes of the partial derivatives
(\ref{partials}) of the strain and of the plastic work with respect to the
logarithms of the model parameters.  The partial derivatives become singular
at the cliff shown in Figure~\ref{epsf}, just as does $\epsilon(f)$ (but not
the plastic work).}
\end{figure}

This failure occurs at a cliff in the $\epsilon(f)$ relation.  Unlike the
case of the cliff found in Section~\ref{ignite}, this solution is physically
meaningful only on one side of the cliff, and at the cliff its behavior is
singular, rather than only rapidly varying (with a finite derivative) as a
function of the control parameter.  The partial derivatives of $\epsilon$
with respect to the logarithms of the model parameters $\epsilon_0$ and
$Y^\prime$ are shown in Figure~\ref{epsderiv}.  These partial derivatives
are singular at the cliff.  However, the location $\epsilon_{sing}$ of the
cliff (\ref{epssing}) and the corresponding values of $f$ and $Y$ are smooth
functions of $\epsilon$ and of $Y^\prime$.
\section{An Indeterminate Model: Ignition in the Presence of Growing
Instability}
In order to simulate an indeterminate model we replace the constant
transport coefficient $D$ (which sets the characteristic scale of the model)
by an effective turbulent transport coefficient $D_m$ for the $m$-th member
of an ensemble of realizations
\begin{equation}
\label{randdiff}
D_m = D \left(1 + d_{max} {\zeta_m \exp{\alpha t} \over 1 + \zeta_m
\exp{\alpha t}}\right),
\end{equation}
where
\begin{equation}
\label{epsm}
\zeta_m \equiv \zeta_0 \sqrt{-2\ln{(1-R_m)}}
\end{equation}
is the initial amplitude of a perturbation, $\alpha$ is its growth rate,
$d_{max} + 1 \gg 1$ is an arbitrary upper bound to $D_m / D$), $\zeta_0
\ll 1/d_{max} \ll 1$, and $R_m$ is a random variable uniformly distributed
in the interval (0,1).  The form (\ref{randdiff}) and distribution 
(\ref{epsm}) are chosen to represent the growth and nonlinear saturation of
an instability, such as a perturbation on a Rayleigh-Taylor unstable
interface, if its initial amplitude is the root-mean-square sum of two
independent Gaussians of unit standard deviation.  This would be expected if
its fastest growing wavelength has contributions from sine and cosine terms
that are independently determined by random fluctuations in the initial
conditions.  Such a disturbance would be expected to increase the effective
diffusivity, by turbulent mixing, of heat and composition over the molecular
diffusivity and thermal conductivity, and (\ref{randdiff}) may represent its
effects on a chemically reacting mixture or a laser fusion ignition capsule.

This process amplifies unknowable and very small variations in initial
conditions to magnitudes that may have macroscopic consequences at later
times.  The most familiar example of such amplification is the
unpredictability of the weather; it is proverbially said (though not
rigorously provable) that the flapping of a butterfly's wings or the waving
of a handkerchief changes the weather a year hence.  In the comparatively
short term these changes grow exponentially, described by a positive
Liapunoff exponent, but at long times they saturate and the weather remains
within finite bounds.

For $\alpha t \lesssim 1$, $d_{max} \zeta_m \exp{\alpha t} \ll 1$ (for
all but an exponentially small fraction of the $R_m$) and $D_m \approx D$.
At later times $D_m \to (d_{max}+1) D$.  We take $d_{max} = 3$ and 
$\zeta_0 = 10^{-4}$; these values are arbitrary, and are chosen only to
illustrate the qualitative features of such a model.  Indeterminacy is
maximized on the upper shoulder of the cliff, whose steepest slope occurs at
$T_0 = 1.70$, so we adopt $T_0 = 1.80$.  For small instability growth rates
the dispersion in the final fraction burned is comparatively small because
the instability does not grow much before exhausting the fuel and the
assumed determinate diffusivity brings burning to an end.  At high growth
rates burning is effectively suppressed (it is sensitive to $D$, so that
even comparatively small $d_{max}$ has a large effect) unless $\zeta_m$
happens to be very small ($R_m$ is close to unity); such cases provide a
``tail'' of larger burnup fractions and maintain a comparatively large
standard deviation, even though most trials fall into a narrow peak at low
burnup fraction.

For any single $m$ the model is determinate because digital random number
generators are determinate, but the ensemble of results represents the
ensemble that would result from an indeterminate choice of uniformly
distributed $R_m$.  The purist might use for the random number generator an
external, genuinely random, seed (such as the digitized voltage measured
across a warm resistor), or a nearly random external seed (such as the
low-order bits of the wall clock time), but this is not necessary in order
to determine the distribution of indeterminate results.

\begin{figure}
\begin{center}
\includegraphics{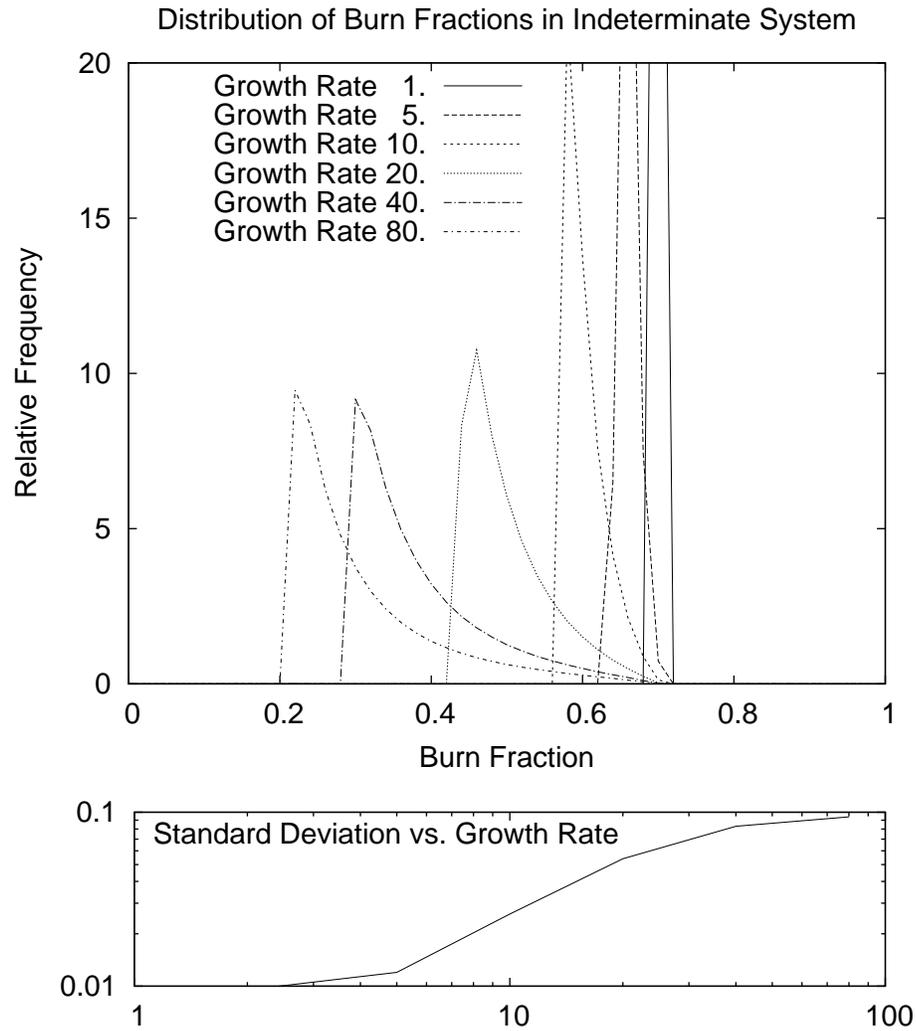}
\end{center}
\caption{Distribution of burnup fractions for indeterminate model,
qualitatively representing the effects of an exponentially growing 
instability.  The distributions are normalized to a uniformly distributed
mean.}
\label{nondetfig}
\end{figure}

We display results for $\alpha = 1, 5, 10, 20, 40, 80$ in Figure
\ref{nondetfig}.  In the determinate model at the assumed $T_0$ burning is
approximately half-completed at a time $r_{char} \approx 0.6$.  For
$\alpha = 10$ at this time the indeterminate multiplier of $D$ is
$\approx 1 + 0.12\zeta_m$; the small random
variance in the reaction rate has a significant effect because of the choice
of parameter values that place the system near the ignition cliff.  For
$\alpha \gtrsim 20$ the indeterminate multiplier closely approaches its
limiting value of $1 + d_{max} \gg 1$ very early in ignition, and burning is
effectively suppressed unless $R_m$ happens (rarely) to be unusually small.

These results should be compared to the burnup fraction of 0.71 for the same
parameters in the determinate model.  Even very small variations in
initial conditions may produce large variations in the final state (burnup
fraction).  For smaller values of $\alpha \lesssim - \ln{(d_{max}\zeta_0}/
t_{char}$ (insufficient to produce large variations in $D_m$ at the
characteristic half-burn time) these variations are amplified to substantial
values by the sensitivity to initial conditions at a cliff.  For $\alpha
\gtrsim - \ln{(d_{max}\zeta_0)}/t_{char}$ the variations in $D_m$ are large,
with correspondingly larger variations in final burnup.  These may approach
(but not very closely because of the details of the model) the physical
limits of zero or complete burnup.

A brittle material is another example of an indeterminate system.  For
stresses below some limiting value, its strain is a linear function of
stress.  Above this limiting value, the stress {\it vs.\/} strain curve
terminates and the material fails abruptly.  There is no cliff and no
indication in the linear curve of failure at higher stress.  Brittle
materials generally have low Weibull moduli and unpredictable and
indeterminate (within a finite but broad range) failure limits.
\section{Discussion}
Cliffs have these properties:
\begin{enumerate}  
\item Performance degrades steeply from its maximum value near a cliff.
This, in itself, need not be unacceptable; in some circumstances even the
reduced performance may meet the designer's needs.
\item Cliffs in design space are regions of uncertain performance.  The
$Y_k$ are sensitive to the (generally poorly known and often large)
uncertainties in the uncontrolled $\{x_i\}$ for the same range of the
design parameters $\{p_j\}$ as those (the cliff) for which the $Y_k$ are
sensitive to the controlled $\{p_j\}$.  This is shown in the first model
problem by the fact that the cliff in Fig.~\ref{cliff}, showing the
sensitivity of energy release to $\{p_j\}$, is found in the same range of
$T_0$ (the only element of $\{p_j\}$ in the model) as the peaks in partial
derivatives in Fig.~\ref{derivs}.  In the second problem an analogous
conclusion may be drawn by comparing the location of the cliff in
Fig.~\ref{epsf} to the sensitivities in Fig.~\ref{epsderiv}.
\item In both problems, the peaks in sensitivity to all the uncertainties
occur for the same values of the design parameters.  This conclusion 
plausibly applies also to unknown uncertainties (``unknown unknowns'') in
real systems, and emphasizes the importance of constraining them.
\item Cliffs are evident in plots of performance as a function of known
control parameters, such as Figs.~\ref{cliff}, \ref{epsf}.  The significance
of the peaks in partial derivatives, such as shown in Figs.~\ref{derivs} and
\ref{epsderiv}, is the demonstration that near these cliffs performance is
sensitive to all parameters of the problem, including those not controllable
by the designer but reflecting intrinsic uncertainties of physical
properties or processes or of initial conditions.
\end{enumerate}

We have illustrated parameter sensitivity near cliffs in a very simple 
model systems that are not chaotic; any particular initial conditions lead
smoothly to a stationary final state, even though that state cannot be
predicted from imprecise knowledge of the initial conditions.  Analogous
phenomena are found in many other systems, such as the behavior of an
elastic column under compression near its Euler buckling threshold, or the
orbit of a spacecraft deflected by a close approach to a planet.

Some classical systems are effectively indeterminate.  An example is the
initiation of a detonation wave in high explosive near its threshold shock
initiation pressure; initiation depends on the unmeasurable
details of voids and heterogeneity, but the final state will be either
nearly complete detonation or failure of more than a small quantity
explosive to detonate.  Additional examples of nonchaotic indeterminacy
include brittle fracture resulting from the growth of microscopic flaws such
as Griffith cracks (Griffith 1920) and the nucleation of phase transitions,
whether homogeneous (resulting from intrinsically unpredictable fluctuations
in thermodynamic equilibrium) or heterogeneous (resulting from the presence
of nucleation sites that are, in practice, uncharacterizable to an accuracy
sufficient to make the system determinate). 

In general, systems in which unquantifiable small initial variations grow
exponentially may be indeterminate over a finite, sometimes wide and
important, range of final states.  Even when the initial state is apparently
well-characterized, the finite accuracy of its characterization leads to 
exponentially growing uncertainty.  This is often realized in the form of
chaos.  Examples include the weather, the growth of hydrodynamic instability
in which the detailed final configuration is not predictable (such as the
location of bubbles and spikes in Rayleigh-Taylor instability), and the
formation of caustics in wave propagation through turbulent media.
\section{Conclusions}
Study of these models has led to two important conclusions:
\begin{enumerate}
\item The sensitivities of the $\{Y_k\}$ to the $\{x_i\}$ have narrow maxima
at performance cliffs in the control parameters $\{p_j\}$.  This result is
not surprising, but it is important.  Large values of
\begin{equation}
\left\vert {\partial Y_k \over \partial \ln{p_j}}\right\vert,
\end{equation}
defining a cliff, identify the values of the $\{p_j\}$ for which the $Y_k$
are {\it also\/} sensitive to uncertainties, often poorly quantified, in the
$\{x_i\}$.  Performance is less reliably predictable near a cliff because
there the effects of uncertainties in the $\{x_i\}$ are magnified.
\item The sensitivity of the $Y_k$ to the $\{x_i\}$ shown in Figures
\ref{derivs} and \ref{epsderiv} peak at the {\it same\/} values of the
$\{x_i\}$ for all the $x_i$ in the model.  This leads to the generalization
that in a general design problem the sensitivity of the $\{Y_k\}$ to all
uncertainties, including ``unknown (or underestimated) unknowns'', has a
narrow maximum for values of the design or control parameters $\{p_j\}$
near a cliff.  Even if the known sensitivities to uncertainties in
the $\{x_i\}$ are small enough that the resulting uncertainties in the $Y_k$
are acceptable, design near a cliff introduces the risk that unknown
uncertainties will have unacceptably large consequences. 
\end{enumerate}

Because the uncertainties near a cliff are proportional to the partial
derivatives \ref{partials} that generally have sharp maxima there, the
ratio $M/U$ of margin to uncertainty (Goodwin \& Juzaitis 2006, Eardley
{\it et al.\/} 2005, Pilch, Trucano \& Helton 2006) may have a sharp
minimum at a cliff.  Prudent design requires a minimum $M/U$ whose value
depends on how well the system is understood (equivalently, with how much
confidence $M$ and $U$ can be calculates).  Optimization of design in the
presence of resource constraints tends to minimize margin, making optimized
designs particularly subject to the increased uncertainty near cliffs.  The
art of engineering design comprises optimizing the trade-offs among these
conflicting requirements.

This qualitative behavior is a general property of nonlinear systems in
which there are two (or more) competing but interacting processes, each
capable of runaway growth.  Depending on the quantitative values of the
parameters, one or the other may dominate.  It is unavoidable that there be
a sharp dividing line between these regimes, which may be though of as a
knife-edge ridge in parameter space, in which performance is a sensitive
function of the values of the parameters.  This dividing line corresponds to
a cliff in a map of performance as a function of the parameters.

Performance is intrinsically sensitive to uncertainty near and on
performance cliffs.  A prudent designer attempts to avoid these regimes,
even if the nominal calculated performance within them is sufficient for his
purposes.

The importance of these conclusions is their robustness.  We have
found them in two very different classes of problems, and in cases in which
there are random as well as non-random initial conditions.  We expect them
to be applicable to a very broad range of engineering designs.

I thank B.~Cheng, F.~J.~Dyson, F.~Graziani, J.~Pedicini and S.~Sterbenz for
discussions.  Work at the Los Alamos National Laboratory performed under the
auspices of the U.~S.~Department of Energy under Contracti
DE-AC-52-06NA25396.  Work at the Lawrence Livermore National Laboratory
performed under the auspices of the U.~S.~Department of Energy under
Contract DE-AC-52-07NA27344.

\end{document}